\documentclass[sigconf]{acmart}
\usepackage{makecell}
\usepackage{threeparttable}
\usepackage{multirow}
\usepackage{subfig}
\usepackage{booktabs}
\AtBeginDocument{%
  }

\copyrightyear{2025}
\acmYear{2025}
\setcopyright{cc}
\setcctype{by-nc-sa}
\acmConference[ICMI '25]{Proceedings of the 27th International Conference on Multimodal Interaction}{October 13--17, 2025}{Canberra, ACT, Australia}
\acmBooktitle{Proceedings of the 27th International Conference on Multimodal Interaction (ICMI '25), October 13--17, 2025, Canberra, ACT, Australia}
\acmDOI{10.1145/3716553.3750785}
\acmISBN{979-8-4007-1499-3/2025/10}
\begin{document}

\title[Analyzing Character Representation in Media Content]{Analyzing Character Representation in Media Content using Multimodal Foundation Model: Effectiveness and Trust}
\author{Evdoxia Taka}
\affiliation{%
  \institution{University of Glasgow}
  \country{UK}
}

\author{Debadyuti Bhattacharya}
\affiliation{%
  \institution{University of Glasgow}
  \country{UK}
}

\author{Joanne Garde-Hansen} 
\affiliation{%
  \institution{University of Leeds}
  \country{UK}
}

\author{Sanjay Sharma} 
\affiliation{%
  \institution{University of Warwick}
  \country{UK}
}

\author{Tanaya Guha}
\affiliation{%
  \institution{University of Glasgow}
  \country{UK}
}

\renewcommand{\shortauthors}{Taka et al.}

\begin{abstract}
Recent advances in AI has made automated analysis of complex media content at scale possible while generating actionable insights regarding character representation along such dimensions as gender and age. Past works focused on quantifying representation from audio/video/text using AI models, but \emph{without} having the audience in the loop. We ask, even if character distribution along demographic dimensions are available, how useful are those to the general public? Do they actually \emph{trust} the numbers generated by AI models? Our work addresses these open questions by proposing a new AI-based character representation tool and performing a thorough user study. Our tool has two components: (i) An analytics extraction model based on the Contrastive Language Image Pretraining (CLIP) foundation model that analyzes visual screen data to quantify character representation across age and gender; (ii) A visualization component effectively designed for presenting the analytics to lay audience. The user study seeks empirical evidence on the \emph{usefulness} and \emph{trustworthiness} of the AI-generated results for carefully chosen movies presented in the form of our visualizations. We found that participants were able to understand the analytics in our visualizations, and deemed the tool `overall useful'. Participants also indicated a need for more detailed visualizations to include more demographic categories and contextual information of the characters. Participants' trust in AI-based gender and age models is seen to be moderate to low, although they were not against the use of AI in this context. Our tool including code, benchmarking, and the user study data can be found at \emph{\url{https://github.com/debadyuti0510/Character-Representation-Media}}.
\end{abstract}

\begin{teaserfigure}
\centering
  \includegraphics[width=0.9\textwidth]{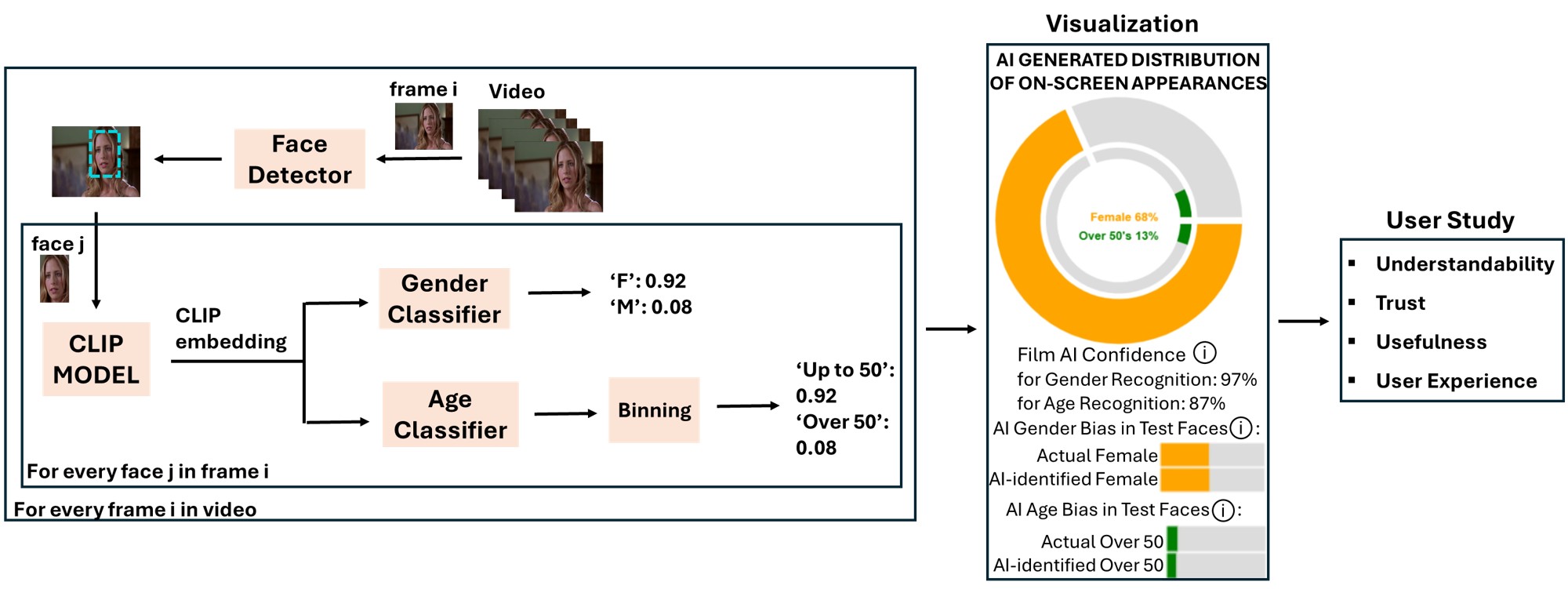}
  \caption{Our tool to analyze the demographic character representation in videos. It detects the faces in the video, generates the face CLIP embeddings and uses them as input to gender and age classifiers. The analytics are visualized at the intersection of gender and age including information about the confidence and bias of the models. We evaluated our tool with human users.}
  \Description{The pipeline for the analysis of the video is presented on the left and the visualization of the analytics on the right at the output of the analysis pipeline. The visualization includes a title stating AI-Generated distribution of on-screen appearances. Below the title there are two nested doughnut rings (the outer is wider than the inner). The outer show the female/male proportion, while the inner the over 50/up to 50 proportions for each gender category. Orange is used for female and green for the over 50 characters. In the middle of the doughnut chart some text states in orange ``Female 68\%'' and below this in green ``Over 50's 13\%''. Below the doughnut chart there is text stating the AI confidence for the gender and age prediction, which is 97\% and 87\%. Below this there are two sets of two bar graphs; one on the left with the top bar graph presenting the proportion of female faces and the bottom graph showing the proportion of AI-recognized female faces in a test dataset. The set of bar graphs on the right, shows the corresponding proportions for the Over 50 category.}
  \label{fig:teaser}
\end{teaserfigure}
\maketitle
\section{Introduction}
How media content is created, produced and consumed have changed rapidly over the last decade. While non-traditional platforms and media forms have become mainstream, AI has simultaneously been integrated into every stage of the media life cycle — from production to audience engagement to consumption. Given the impact of media on our society, it is crucial to ask \emph{how does media represent and reflect society along human dimensions}, such as gender, age, ethnicity, ability, profession, and socioeconomic status? \cite{Somandepalli2021}.

The problem of quantifying representation in media has been addressed by media researchers using traditional methods from screen research that involves hand counting \cite{Smith2019, CMS2019}. On the computational front, this is achieved using video/audio/language processing combined with machine learning (ML) models \cite{Guha2015, Mazieres2021, Somandepalli2021, bamman2024}. Such AI-based content analyzes have been shown to uncover hidden biases in character portrayal in Hollywood movies, such as gendered gaps in screen and speaking time \cite{Guha2015,ramakrishna2017linguistic}. The methodologies used and developed in the context of character representation have primarily focused on analysing the visual streams and faces \cite{bamman2024, Mazieres2021}. The general pipeline includes face detection followed by unsupervised or semi-supervised character discovery and classification of the extracted faces along various dimensions \cite{Guha2015,Kulshreshtha2018, bamman2024}. A major challenge however is the variability and the long form of media. The characters' appearances are diverse and may change widely in long form media (e.g., movies), which makes generalization difficult for ML models. This is where foundation models are expected to play an important role with their superior ability to generalize across diverse content. This current paper demonstrates the effective use of the multimodal foundation model, Contrastive Language-Image Pretraining (CLIP) \cite{DBLP:journals/corr/abs-2103-00020}, to address the task of quantifying character representation in media content. 

Thanks to the recent advances in AI, it is now possible to analyze media data with reasonable accuracy at scale \cite{bamman2024} and generate actionable insights regarding character representation. However, we note that all the work above extracted statistics and analytics from movies \emph{without} having the audience in the loop. The audience still has little information about the diversity in the media content they consume; nor do they have the specialized knowledge or tools required to gather such statistics. More importantly, even if the audience has access to representation statistics, how useful are those numbers to them? Do they actually \emph{trust} these numbers generated by AI models? Our work attempts to answer these questions. 

In this paper, we explore the effectiveness of the vision-language foundation model, CLIP \cite{DBLP:journals/corr/abs-2103-00020}, for analyzing the demographic dimensions (gender and age) of the characters' faces that appear on screen. Following CLIP's zero-shot paradigm, we train two classifiers to predict \textit{perceived} gender and age; the models use CLIP embeddings of the face images detected in long videos. Our gender and age detectors are trained and benchmarked on a large publicly available dataset before they are used to extract analytics from the full-length movies. Benchmarking results show that our gender and age models perform close to the state-of-the-art. We extract analytics from three full length movies and create visualizations with the lay audience as target users. The design of the visualizations is informed by graphs commonly used to present demographic data to the public \cite{GoogleArticle,GDI_chinareport,GDI_indiareport,Smith2019,GDI_stem}. Next, we provide empirical evidence on the usefulness and trustworthiness of the AI-generated results presented in form of our visualizations through a user study.  
To summarize, our main contributions are:
\begin{itemize}
    \item Building an open-source tool based on CLIP to analyze visual screen data for understanding character representation across dimensions of age and gender.
    \item Designing a suitable visualization tool to present such analytics to lay audience.
    \item Presenting evidence from human participants through a rigorous user study on the trustworthiness and usefulness of the analytics generated by the frontier AI tool.
\end{itemize}
\vspace{-1mm}
\section{Related work}
Traditionally, automated analysis of media content has been focused on addressing the needs of organizing, indexing and navigating through large media data corpora. More recently, computing effort is being driven towards generating insights and human-centred analytics from large volumes of media content across audio, video and text \cite{bamman2024,Mazieres2021,Somandepalli2021}. Unimodal approaches such as those of \citet{bamman2024} and \citet{Mazieres2021} automatically quantify character representation in media through analysing characters' faces. These works used the visual data stream with classical ML models to analyze characters' faces. \citet{bamman2024} also studies on ethnic diversity, but this was done through human annotations instead of ML. \citet{ramakrishna2017linguistic} combined linguistic analysis with ML to reveal systematic patterns in character portrayals, such as ethnic minorities not interacting with other characters. 

Existing work on multimodal approaches leverage the inherent multimodal nature of media data to analyze content for understanding character representation. The earliest work include that of \citet{Guha2015}, which developed a multimodal system involving speaker diarization (i.e., who speaks when), face detection and gender identification to measure speaking time and screen time of male and female actors. Researchers have also used cross-domain information to enhance basic tasks such as gender detection or active speaker detection in movies \cite{hebbar2018improving,sharma2019toward}. Recent advancements in multimodal foundation models and large language models (LLMs) have led to a shift of the content analysis towards this direction. \citet{Gan2023} developed a CLIP-based method leveraging text and image analysis for apparent personality perception. \citet{Cerit2025} developed the \emph{Media Content Atlas}, which is a pipeline to explore media content in the multidimensional media space using multimodal LLMs. 

In this work, we use the CLIP multimodal foundation model to quantify the gender and age representativeness in long videos (films). Such an analysis can reveal insights and reveal hidden patterns and biases across demographics \cite{Guha2015,bamman2024,Mazieres2021}, which enable analysis at scale. We specifically focus on the representation of ageing women (\emph{women over 50}) in this work. Various studies in the past have identified a gap in representation for this demographic group \cite{CAB_2023,GDI_2021,Swift2020}. These analytics are communicated to other researchers, the public, or media industry stakeholders through text and visualizations (e.g. bar graphs, pie charts etc.). Many examples of such presentations can be found in publicly available reports from organizations like the \emph{Geena Davis Institute} and \emph{Centre for Ageing Better} \cite{CAB_2023,GDI_2021,Swift2020,GDI_chinareport,GDI_indiareport,Smith2019,GDI_stem}, or corresponding articles in the news/research journals. Nonetheless, there is no available evidence, to the best of our knowledge, what analytics might be relevant and useful to the audience, and how to best visualize these to create a meaningful presentation and trustworthiness.
\section{Methodology}
 Fig.~\ref{fig:teaser} presents an overview of the video analysis pipeline along with the visualization tool that we developed. Below we explain the methodology used to develop this tool.

\subsection{CLIP-based video analysis}
\vspace{1mm}
\noindent\textbf{Face detection and CLIP embeddings}.
Our video analysis pipeline, like several past works, relies on analyzing only the characters' faces that are seen in the videos. We use the DeepFace's \cite{serengil2020lightface} \emph{Single Shot Face Detector} \cite{8237292} as our face detector. For each detected face we use the \emph{Huggingface} transformers package \cite{DBLP:journals/corr/abs-1910-03771} to generate their CLIP embeddings. The CLIP model itself is trained on the WebImageText dataset \cite{DBLP:journals/corr/abs-2103-00020}. 

\vspace{1mm}
\noindent\textbf{Perceived gender and age detector}.
We train two Logistic Regression (LR) models using the CLIP embeddings as inputs to predict the perceived gender and age of each detected face: $ P(y = k \mid \mathbf{x}) = \frac{\exp(\mathbf{w}_k^\top \mathbf{x} + b_k)}{\sum_{j=1}^{K} \exp(\mathbf{w}_j^\top \mathbf{x} + b_j)}
    \label{eq: lr_clip}
$, where \( \mathbf{x} \) is the face CLIP image embedding, \( P(y = k \mid \mathbf{x}) \) is the probability that the prediction \( y \) belongs to class \( k \) given \( \mathbf{x} \),  \( \mathbf{w}_k \) is the weight vector corresponding to class \( k \), \( b_k \) is the bias term corresponding to class \( k \), \( K \) is the total number of classes. The numerator represents the exponential function applied to the linear combination for class \( k \) and the denominator serves as a normalizing factor that divides the numerator by the sum of exponents for all classes. 

The classes for the perceived gender are Female/Male and perceived age groups are: 0-2, 3-9, 10-19, 20-29, 30-39, 40-49, 50-59, 60-69 and 70+. These categories given by the FairFace dataset \cite{DBLP:journals/corr/abs-1908-04913} were used for training the LR classifiers. We choose this dataset because it was designed to provide a diverse and balanced distribution across age groups, gender, and race labels across different demographic groups. Labels for perceived age and gender in this dataset are expected to enable fairer and more generalizable classification performance. 

\vspace{1mm}
\par\noindent\textbf{Benchmarking.}
We first benchmarked the LR models using as a baseline the CLIP ZS model \cite{DBLP:journals/corr/abs-2103-00020} and the deployed gender and age detectors of the DeepFace library \cite{serengil2020lightface} (these models are trained on the IMDB-WIKI dataset \cite{Rothe-IJCV-2018}). For the validation, we used the validation set of the FairFace dataset \cite{DBLP:journals/corr/abs-1908-04913}. The deepface age model predicts the ages in the set \{0,1,...,100\} and we mapped each prediction to the considered age groups. The text we used to describe the classes we considered in the case of the CLIP Zero-Shot model (CLIP ZS) was ``the face of a \{man,woman\}'' and ``A person in the \{x\} age group'' for gender and age, respectively.
Table~\ref{table:gender_age_table} shows that both CLIP ZS and CLIP + LR outperform the DeepFace deployed model for the prediction of gender and age in the FairFace dataset. The performance of these models is comparable to the state-of-the-art performance in the FairFace dataset \cite{Kuprashevich2024} with accuracy \textbf{97.5}\% and \textbf{62.28}\% for gender and age. 

\begin{table}[!h]
\caption{Gender and age prediction performance of the CLIP-based models on the FairFace dataset.}
\vspace{-3mm}
\resizebox{0.8\linewidth}{!}{%
\begin{tabular}{@{}lcc@{}}
\toprule
\bf Model &
  \bf Accuracy (\%) &
 \bf F1-Score (Macro) \\ 

\midrule
DeepFace Gender & 77.93  
& 0.77 \\
CLIP ZS Gender & 95.06  
& 0.95 \\
CLIP+LR Gender
& \textbf{96.16}  
& \textbf{0.96} \\ 
\midrule
DeepFace Age 
& 26.63 
& 0.15 \\
CLIP ZS Age 
& 39.84  
& 0.37 \\
CLIP+LR Age 
& \textbf{60.13} 
& \textbf{0.57} \\
\hline
\end{tabular}
}
\label{table:gender_age_table}
\end{table}

\vspace{2mm}
\par\noindent\textbf{Results on full-length films.} We chose three full-length films to generate the gender and age analytics to be used for visualization and subsequent user study. These include Mamma Mia! The Movie (2008), The Best Exotic Marigold Hotel (2011), and Black Panther (2018). 
The films were selected to include diverse distributions of on-screen appearances for gender and age. The films and analysis results are shown in Table~\ref{table:films_results}.
\begin{table}[!h]
\caption{Gender and age distribution using our tool}
\vspace{-3mm}
    \centering
    \resizebox{\linewidth}{!}{%
    \begin{tabular}{l p{4cm} c c c}
    \toprule    
    \bf ID & \bf Film     & \bf \#Faces & \bf Female (\%) & \bf Age>50 (\%) \\
    \hline
    1& Mamma Mia (2008) & 6,841 & 68.29 & 12.52\\
    2& Marigold Hotel (2011)& 2,723 & 66.66 & 46.29\\
    3&Black Panther (2018) & 2,677 & 27.22 & 8.88\\
    \bottomrule
    \end{tabular}
}
    \label{table:films_results}
\end{table}
\subsection{The visualization tool}
\label{sec:visualization}
Fig.~\ref{fig:teaser} presents our visualization created to show the gender and age representation \emph{at their intersection} i.e., for women characters over 50. Our visualization is inspired by the doughnut charts commonly used to present demographic data to the public \cite{GoogleArticle,GDI_chinareport,GDI_indiareport,Smith2019,GDI_stem}. Such graphs are widely used in the news, reports, etc., and the general audience is expected to be familiar with them. We alter the standard design of the doughnut chart to embed a second ring within it with the purpose to present the distribution of appearances at the intersection of gender and age. The visualization is enhanced with popups on hover to show the exact percentages corresponding to each demographic group and explanations for what the AI prediction confidence and bias means. 

In the case of age, we map the age predictions to the set \{`Up to 50',`Over 50'\}. We assign each detected face to the appropriate group in this set depending on which of the following sums is greater: the sum of confidences for all age groups with ages>50 and the corresponding sum for ages<50.

In existing studies \cite{CAB_2023,GDI_2021,Swift2020,GDI_chinareport,GDI_indiareport,Smith2019,GDI_stem}, the demographic representation in media is usually measured at the level of characters, e.g,. number of female/male characters appearing or speaking. We measure this at face-level indicating overall on-screen appearances, e.g., the number of detected faces predicted as female or not to avoid the extra error from the character discovery algorithm. We also incorporate information about our models' confidence and bias, following the paradigm of Responsible AI. We show the confidence score for the ML models while making the predictions below the graph, and the model's bias for the prediction of gender and age as bar graphs at the bottom. For the confidence score, we calculate the average confidence of the dominant gender/age predictions across all faces. For the bias, we show the actual and predicted percentages of each category in the validation set of the FairFace dataset.
\section{User study}
We conduct a user study to assess the understandability of the presented information, the usefulness of our tool, and the trust in our model for the general audience. In particular, we concentrate on the following research questions:
\begin{itemize}
    \item[RQ1] Do people understand the character representation information shown to them?
    \item[RQ2] To what extent do people trust the AI-generated results?
    \item[RQ3] How useful is our tool for the audience?
    \item[RQ4] What is the user experience when using our tool?  
\end{itemize}
\subsection{Procedure}
\begin{figure}[tb]
    \centering
    \includegraphics[width=\linewidth]{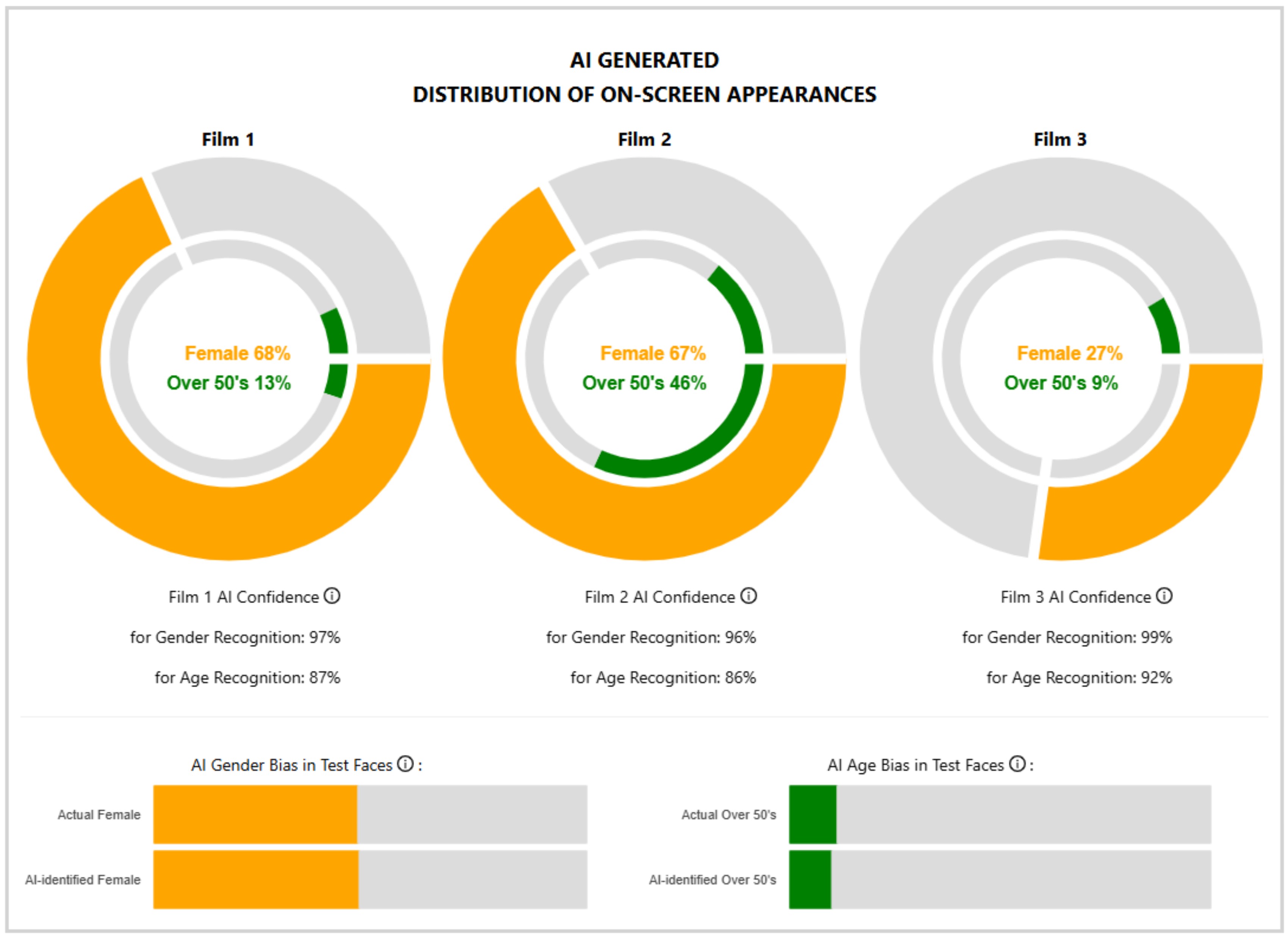}
    \vspace{-3mm}
    \caption{Charts shown to participants with character representation statistics of the three films in Table~\ref{table:films_results} while withholding the film names.}
    \label{fig:studygraphs}
    \Description{}
\end{figure}

We generate the visualizations (see Fig.~\ref{fig:studygraphs}) for the 3 films named in Table~\ref{table:films_results}  and use them to evaluate our tool with human users. We do not disclose the names of the films to participants to avoid any bias due to prior knowledge or perception. 
We recruited 30 participants through University of Glasgow mailing lists and social media channels. The participants spent 30 minutes at max and they were compensated for their time. The user study was approved by the Ethics Committee of University of Glasgow. After the participants provide consent, they can proceed to the study questions. 
\begin{figure*}[!t]
    \centering
    \includegraphics[width=0.85\linewidth]{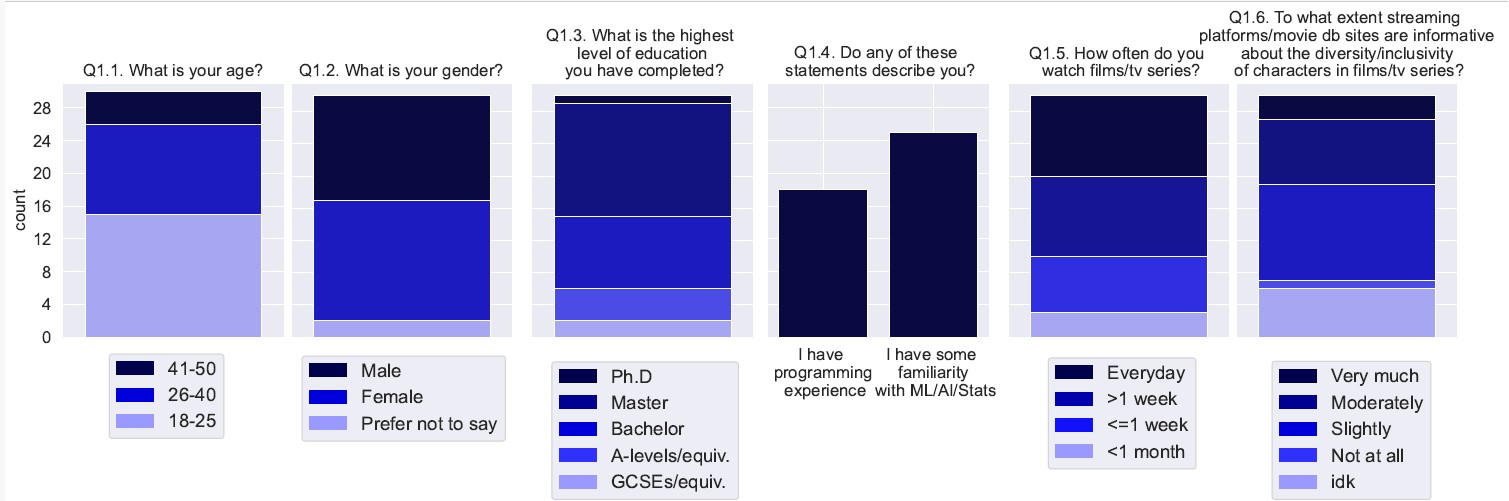}
    \caption{\textbf{Part 1}: pre-questionnaire \& distribution of responses.}
    \label{fig:prequestionnaire}
    \Description{}
\end{figure*}
\begin{table*}[!h]
\caption{\textbf{Part 2-3}: overview of questions and mapping to research questions (RQs). (\emph{idk}: ``I don't know'' )}
\vspace{-3mm}
\resizebox{\linewidth}{!}{%
\begin{tabular}{@{}lp{14cm}p{8cm}p{3cm}@{}}
\toprule
Code &
  Question &
  Response Options &RQ\\ 
\midrule
Q2.1 & What does the following information about Film 3 mean? Female 27\% & 27\% of the \textbullet characters appearing in the film are female \textbullet cast of the film are female \textbullet total number of on-screen character appearances in the film belong to female characters \textbullet time characters speak in the film the speech belongs to female characters \textbullet idk &RQ1 - understandability\\
Q2.2 & Which of the following categories of characters appears the most on screen in Film 1 according to the AI? & \textbullet Female \textbullet Male \textbullet idk &RQ1 - understandability\\
Q2.3 & Which of the following categories of characters appears the most on screen in Film 2 according to the AI? & \textbullet Characters <= 50 years \textbullet Characters > 50 years \textbullet idk &RQ1 - understandability\\
Q2.4 & Which of the following categories of characters appears the least on screen in Film 3 according to the AI? & \textbullet Female <= 50 \textbullet Female > 50 \textbullet Male <= 50 \textbullet Male > 50 \textbullet idk &RQ1 - understandability\\
Q2.5 & Which film proportionately presents a fair representation of characters over 50? & \textbullet Film 1 \textbullet Film 2 \textbullet Film 3 \textbullet idk &RQ1 - understandability\\
Q2.6 & Which film proportionately favours the representation of Female characters over 50? & \textbullet Film 1 \textbullet Film 2 \textbullet Film 3 \textbullet idk &RQ1 - understandability\\
Q2.7 & For which film the AI is more confident when recognizing the age of the appeared faces in the film? & \textbullet Film 1 \textbullet Film 2 \textbullet Film 3 \textbullet idk &RQ1 - understandability\\
Q2.8 &  Which of the following categories do you think the AI is likely to underdetect - i.e., recognize fewer instances than are actually present? (E.g., if 10 women appear, the AI might only detect 7.)  Please assume > 1\% bias. & \textbullet Female \textbullet Male \textbullet Up to 50 \textbullet Over 50 \textbullet idk &RQ1 - understandability\\
Q2.9 & How confident are you in the gender predictions the AI makes for the appeared faces in films? & \textbullet not at all \textbullet slightly \textbullet moderately \textbullet very \textbullet extremely confident &RQ2 - trust\\
Q2.10 & How confident are you in the age predictions the AI makes for the appeared faces in films? & \textbullet not at all \textbullet slightly \textbullet moderately \textbullet very \textbullet extremely confident &RQ2 - trust\\
\hline
Q3.1 & How useful do you think the pie charts in this context are? & \textbullet not at all \textbullet slightly \textbullet moderately \textbullet very \textbullet extremely useful &RQ3 - usefulness\\
Q3.2 &  How useful do you think is presenting the on-screen appearances distrib. at the intersection of gender and age? & \textbullet not at all \textbullet slightly \textbullet moderately \textbullet very \textbullet extremely useful &RQ3 - usefulness\\
Q3.3 &  How useful do you think the use of AI in this context is? & \textbullet not at all \textbullet slightly \textbullet moderately \textbullet very \textbullet extremely useful &RQ3 - usefulness\\
Q3.4 &  How useful do you think the inform. about the AI conf. levels for the recogn. of gender and age in films are? & \textbullet not at all \textbullet slightly \textbullet moderately \textbullet very \textbullet extremely useful &RQ3 - usefulness\\
Q3.5 &  How useful do you think the information about the AI bias levels in the recognition of gender and age are? & \textbullet not at all \textbullet slightly \textbullet moderately \textbullet very \textbullet extremely useful &RQ3 - usefulness\\
Q3.6 &  How cognitively demanding it was to read and understand the AI-generated demographic analytics of the films based on this presentation? & \textbullet not at all \textbullet slightly \textbullet moderately \textbullet very \textbullet extremely demanding &RQ4 - user experience\\
Q3.7 &  How much insecure, discouraged, irritated, stressed, and annoyed were you when looking at this presentation of the AI-generated demographic analytics of the films? & \textbullet not at all \textbullet slightly \textbullet moderately \textbullet very \textbullet extremely &RQ4 - user experience\\
Q3.8 &  Please rate the AI tool for analysing the character demographics. & 5 star scale &RQ4 - user experience\\
Q3.9 &  How much do you like using this tool? & \textbullet not at all \textbullet a little \textbullet somewhat \textbullet a lot \textbullet completely  &RQ4 - user experience\\
Q3.10 &  Please tell us a bit more about what you thought about this tool. What did you like or dislike? Was there anything else you wanted to see but wasn’t able to? (Please describe your answer in as much detail as possible.) & open-ended &RQ4 - user experience\\
Q3.11 &  If you had access to a tool like this one (a visual presentation of AI-generated analytics for gender and age in screen media), how would you use it? & open-ended &RQ3 - usefulness\\
\hline
\end{tabular}
}
\label{table:questions}
\end{table*}

\begin{figure*}[!h]
    \centering
    \subfloat[]{\includegraphics[width=0.35\linewidth]{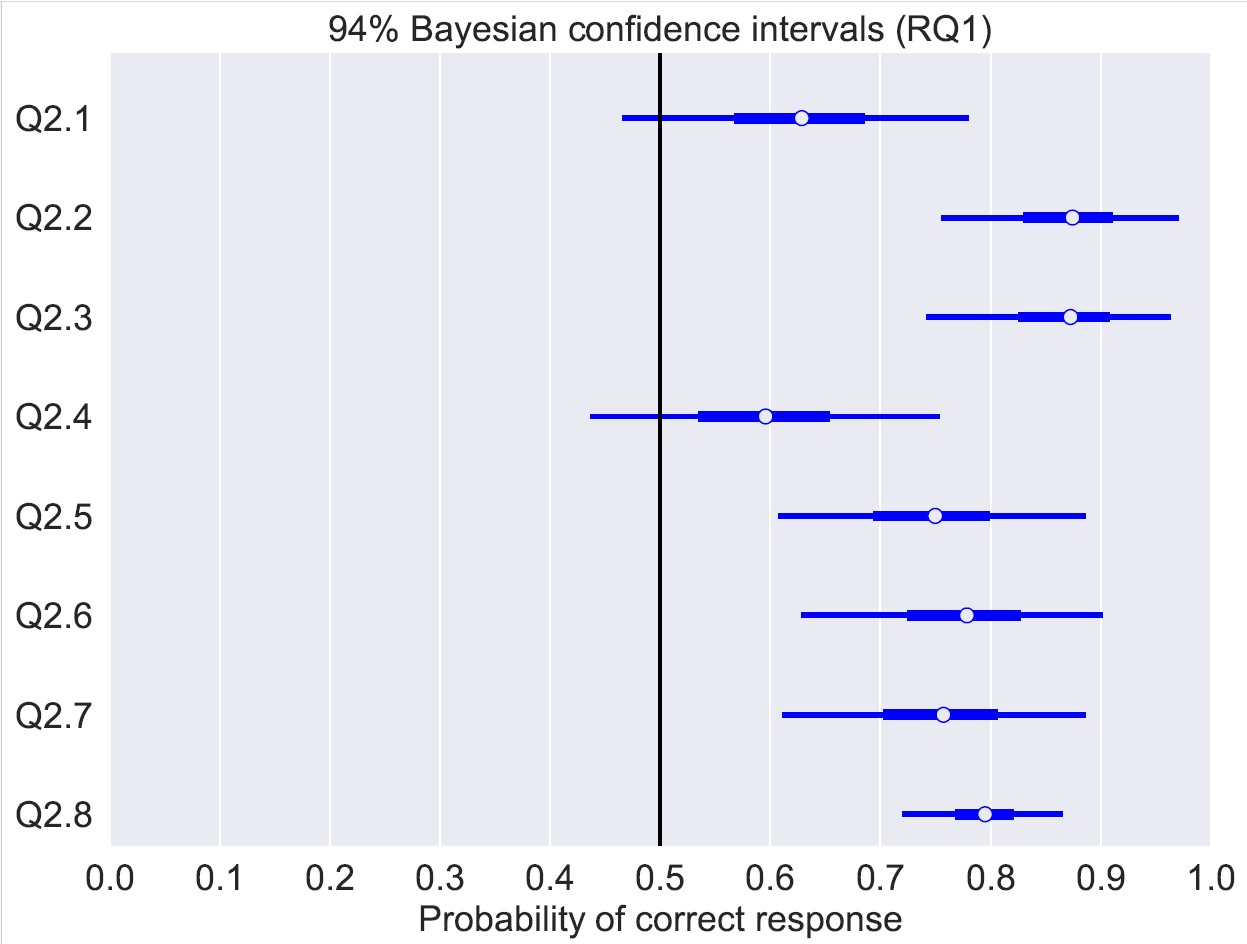}
    \label{fig:understandability}}\quad\quad\quad
    \subfloat[]{\includegraphics[width=0.35\linewidth]{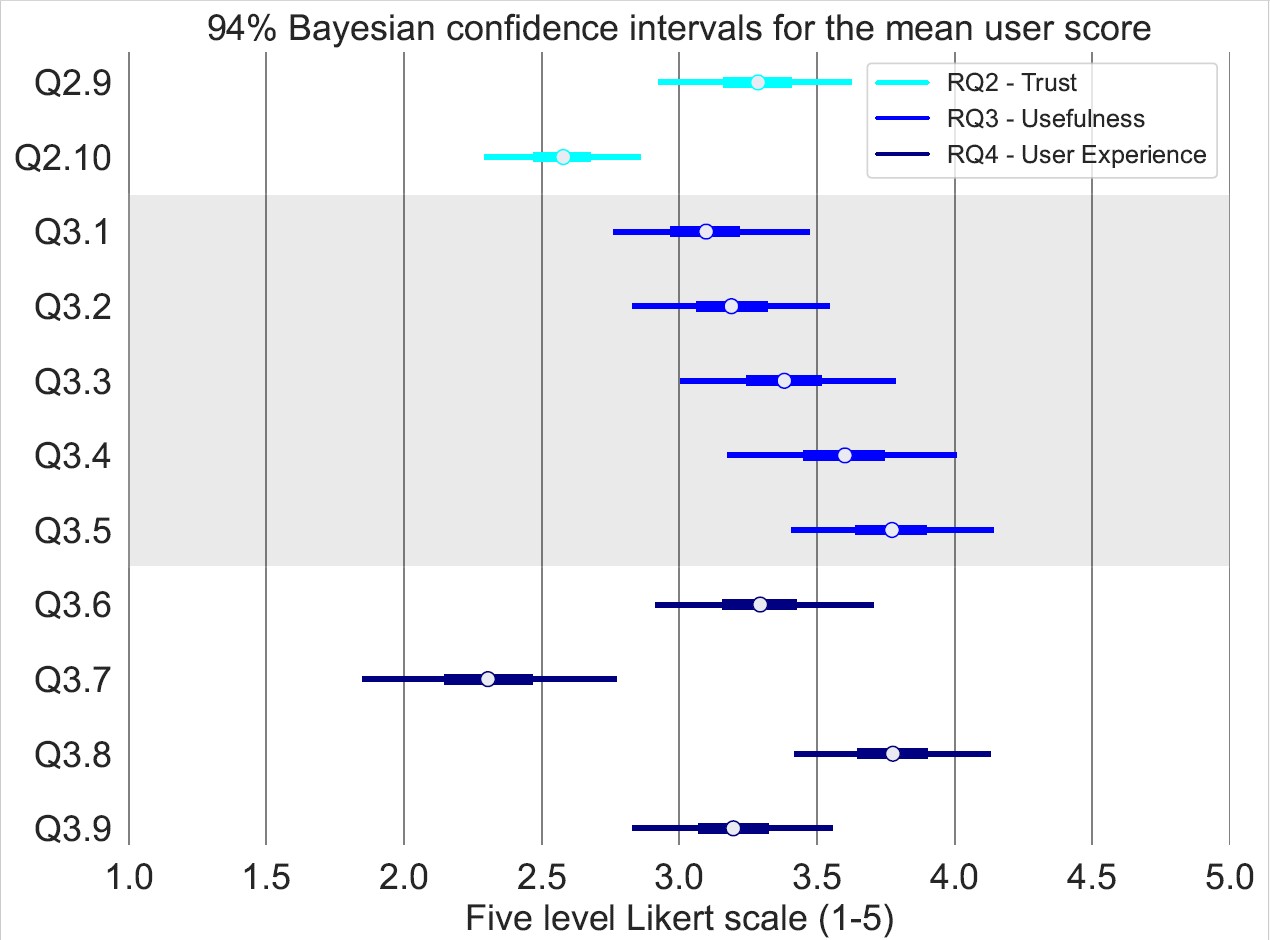}\label{fig:liketscores}}
    \caption{Part 2-3: 94\% Bayesian CIs for (a) the probability of correct response in the information \emph{understandability} questions and (b) the average user scoring of their \emph{trust} in our models, the \emph{usefulness} of our tool, and their \emph{user experience}.}    
    \Description{}
\end{figure*}
The user study comprise three parts: \emph{Part 1} captures participant demographics (questions noted as Q1.x). Fig.~\ref{fig:prequestionnaire} presents the distribution of Part 1 responses. \emph{Part 2} provides a link to the film visualizations (Fig.~\ref{fig:studygraphs}) and tests participants' understanding of the presented information and trust in the ML models (questions noted as Q2.x). \emph{Part 3} investigates the tool usefulness and the user experience (questions noted as Q3.x). Table~\ref{table:questions} presents an overview of the question in Parts 2 and 3 mapped to the RQs.
\subsection{Analysis} 
We conduct statistical Bayesian inference for the closed-ended questions similarly to \citet{Taka2024} (Bayesian analysis can provide useful inferences even for small sample sizes). These questions can be categorized in two groups for the purpose of analysis. 

First, questions providing a list of options to select the correct response(s) (Q2.1-8 - understandability). We measure the accuracy of each participant in every question. Participants’ responses are transformed into a binary representation with 0 indicating a wrong and 1 a correct selection. Participants that selected the ``I don't know'' (\emph{idk}) option were excluded. This option was included to avoid noise by random responses. Very few people picked this option (and not in all questions); Q2.5/6:  1, Q2.2:  2, Q2.3:  3, Q2.8:  4.  Participants’ transformed responses in Q2.1-7 (single selection was allowed) consist of a single digit. In Q2.8 (multiple selections were allowed) this number was 4 (excluding the \emph{idk} option). Participants’ performance in each question is calculated as the number of digit 1 occurrences in their response. We assume a Bernoulli/binomial likelihood for the number of correct responses (depending on whether the single or multiple selections were allowed) with a Beta prior for the probability of a correct answer. Then, we estimate the posterior distribution of this probability as a measure of the users' understanding of the presented information.  

Second, questions asking for a rating in a Likert scale (Q2.9-10 - trust, Q3.1-5 - usefulness, Q3.6-9 - user experience). We assume a normal likelihood for the ratings with a normal prior for the mean and a half-normal prior for the standard deviation of them. We estimate the posterior distribution of the mean rating as a measure of the user reported trust, usefulness, and user experience.
\subsection{Results}
Fig.~\ref{fig:understandability} presents the Bayesian confidence intervals (CIs) for the probability of participants giving correct responses to questions Q2.1-8 based on the posterior distribution. A probability of $0.5$ indicates random responses. CIs that are on the right of the reference line of $0.5$ and do not intersect with it indicate the probability of providing accurate responses. The probability of correct response increases with the distance of the CI from the reference line on the right side of it. Fig.~\ref{fig:liketscores} presents the (Bayesian) CIs for the mean participant ratings in questions Q2.9-10 and Q3.1-9.

\vspace{1mm}
\noindent\textbf{Information understandability (RQ1)}.
Based on our results, the evidence on the ability of the participants to understand the character distribution the doughnut charts' present (cast, characters, on-screen appearances etc.) is not strong (Q2.1).The CI intersects with the reference line and is quite wide spanning between probability 0.47 and 0.78 (mean=0.62). Although the title of the visualizations clearly states that it's the distribution of on-screen appearances  presented (Fig.~\ref{fig:studygraphs}), only 19 out of 30 participants noted this correctly, while 7 participants assumed the distribution refers to the cast of the film, and 4 to the number of 
characters appearing in the films.  
Our results suggest that participants are able to distinguish gender and age information. The participants can identify which categories of appearances between Female/Male or Over/Up to 50 prevail in the films. The CIs for the corresponding questions (Q2.2-3) indicate a high probability of correct responses, above $0.75$ with mean probability $0.87$ and $0.86$, respectively. The design of the visualization has been crucial here: the required information to respond to these questions is clearly stated in the center of the doughnut charts, and is indicated by the colored wedges. This argument is further supported by the results in Q2.5. Participants are able to correctly identify the film with the fair representation of the category `Over 50', albeit the wider confidence interval for the probability of correct response (it spans between 0.6 and 0.89, mean=0.75).

We also investigated participants' ability to interpret the information presented at the intersection of gender and age. Responses to Q2.4 were ambiguous, and did not provide us with reliable evidence. This question asked which category at the intersection of gender and age appears the least in Film 3, while `Female Over 50' does not appear at all and `Male Over 50' appears the least. We considered the first as a correct response, but the second could be deemed correct, as well. In action, 18 participants selected the first, 7 selected the second, 3 selected the `Male Up to 50' and 2 the `Female Up to 50'. The CI for Q2.4 intersects with the reference line and is quite wide (it spans between 0.43 and 0.75, mean=0.59). 

Nevertheless, Q2.6 provides a good evidence for this. Participants correctly identify the film that proportionately favors the representation of the category `Female Over 50', albeit the wider CI for the probability of correct response (it spans between 0.64 and 0.91, mean=0.78). The design of the visualization about this point is more complex: the inner circle of the doughnut charts presented the percentage of representation for the groups at the intersection of gender and age with appropriate splits in wedges. The grayed wedges had to be considered here. Interactive pop-ups showed the label of each wedge (colored or not) and the associated percentage.

Finally, our results suggest that participants are able to identify the film with the greatest AI confidence for age prediction (Q2.7) and the category that the AI could likely under-detect (Q2.8). We found a quite wide CI for Q2.7 with probability between 0.61 and 0.89 (mean=0.75), but a much shorter CI for Q2.8 between 0.72 and 0.86 (mean=0.79). The presentation of this information was quite straightforward. The AI prediction confidence was provided in text and the AI bias in bar graphs below the doughnut charts.

\vspace{1mm}
\noindent\textbf{Trust in our models (RQ2)}.
The CI for the mean self-reported participant confidence in the AI predictions for \emph{gender} (Q2.9) has a mean value slightly above the moderate confidence level (3), 3.29, and spans between 2.93 and 3.64, and \emph{age} (Q2.10) has mean value 2.58, and spans between 2.27 and 2.85. This implies that participants moderately trust our gender model, while they only have slight confidence in the predictions of our age model.\\

\vspace{-2mm}
\noindent\textbf{Usefulness of our tool (RQ3)}.
\vspace{-1.4mm}
\subsubsection*{Doughnut charts (Q3.1)} The CI for the usefulness of doughnut charts in this context spans between 2.75 and 3.44 (mean=3.09), meaning they were deemed moderately useful.
\vspace{-1.3mm}
\subsubsection*{On-screen appearances distribution at the intersection of gender and age (Q3.2)} Similarly, the presentation of the on-screen appearances distribution at the intersection of gender and age was deemed moderately useful with a CI spanning between 2.83 and 3.55 (mean=3.19). 
\subsubsection*{Use of AI (Q3.3)} The CI for the usefulness of AI in this context spans between 2.98 and 3.77 (mean=3.39). This implies that participants admit some usefulness of the AI in this case and do not reject it.
\subsubsection*{AI Confidence (Q3.4)} The CI for the usefulness of the AI confidence information spans between 3.2 and 4.03 (mean=3.61). Participants seem to be willing to know about this information. 
\subsubsection*{AI Bias (Q3.5)} The CI for the usefulness of the AI bias information spans between 3.39 and 4.14 (mean=3.77). Participants seem to deem this information rather useful, as well.
\subsubsection*{How would you use this tool? (Q3.11)}
The majority of participants (21) stated some potential applicability/use of our tool. 8 of them clearly referred to some applicability within a film context (e.g., to analyze temporal trends over the decades in films from different countries, cultures, languages, regions, compare across different genres and award winners, analyze how gender and age affect the film's popularity in audience, be implemented into movie review websites). 6 participants thought that this tool would be interesting for doing research. 6 people also mentioned an applicability outside the narrow context of films (digital marketing and media promotion, deployment on social media/content platforms (YouTube, TikTok) to examine potential biases in their recommender systems, to identify the gender of patients to determine their age- and gender-based diet plan in a hospital, for gender recognition of detected faces in CCTV videos, as an AI human identification tool).

3 participants mentioned that they would be interested in seeing more demographics like ethnicity/race in this tool: $\bullet$ ``Develop it further to recognize race, language etc.'' $\bullet$ ``I would be interested to see how accurate AI can analyze the gender and age of people from different ethnic background.'' $\bullet$ ``To check for cultural bias''.

8 participants mentioned that they would (probably) \emph{not} use this tool, and another 2 were not sure about its applicability. 2 of them implied that the counts of appearances is not enough to account for how characters are presented: $\bullet$ ``Representation is very important, but I'm not convinced that the best way to get there is through numerical comparisons.'' $\bullet$ ``There are so many ways this data could be skewed and just having data on gender and age doesn't mean a film reflects positively on the demographics represented. For example, movies like Seven Brides for Seven Brothers would have high instances of women on screen, but that doesn't mean the depictions of women are positive, so to me, a tool like this has absolutely no practical use without further detail provided. Or like in horror movies--yes, there may be many women on screen, but are they all there because they're being tortured and murdered? A tool like this ignores the context in which people appear on screen and would end up generating a lot of false positives in terms of contemporary metrics of representation--it's not enough to just count how many women appear on screen anymore. We need to consider *how* they're depicted.''. 

A participant mentioned that would use such a tool to study ``shifts in representation over time. Expand it to identify shades of characters (e.g. hero/villain, positive/negative roles, victim/perpetrator etc.)''. A participant expressed skepticism about the use of AI. 4 people stated they have no interest in the presented information with 2 of them mentioning they are primarily interested in the film cast.

\vspace{2mm}
\noindent\textbf{User experience (RQ4)}.
\vspace{-2mm}
\subsubsection*{Cognitive Load (Q3.6)} The cognitive load for reading and understanding the presented information was deemed moderate with a CI spanning between 2.89 and 3.69 (mean=3.28).
\vspace{-1mm}
\subsubsection*{Frustration (Q3.7)} Participants' frustration by our visualization was deemed low with a CI spanning between 1.86 and 2.79 (mean=2.31).

\subsubsection*{Tool Rating (Q3.8)} Participants' rating of our tool has a CI spanning between 3.42 and 4.12 (mean=3.77). Participants seem to provide an above average rating to our tool.

\subsubsection*{Liking of using our tool (Q3.9)} Participants' rating of our tool has a CI spanning between 2.83 and 3.54 (mean=3.19). Participants seem to moderately like using our tool.

\subsubsection*{Participants' thoughts about our tool (Q3.10)} We noted that 15 participants made positive comments about (elements of) the presentation or tool often mentioning that they either liked it or found it useful. Examples below:\\
$\bullet$``The tool seems useful, but only if I want to filter movies based on gender''
$\bullet$ ``I like that the parts that are not relevant to subject [...] are greyed out.'' 
$\bullet$``I like the visualization idea a lot''
$\bullet$``Information was presented in a very neat and clean manner'' $\bullet$ ``I think it is a cool tool, and it's impressive how accurate it seems to be'' $\bullet$ ``The graphs were easy to understand'' $\bullet$ ``I like the te[c]h data is presented with pie charts'' $\bullet$ ``This tool was ok-ish I guess. Good observations to be honest.'' $\bullet$ ``I liked the pop ups next to the confidence and bias headings, that was needed.'' $\bullet$ ``I also really liked the info icon for understanding some of the terms'' $\bullet$ ``It was extremely useful to see the comparative values for actual and predicted.'' $\bullet$ ``I liked the figures below the pie chart'' $\bullet$ ``the hover features definitely made it a lot easier to interpret so that was helpful''.

Ten participants found that the presentation was confusing or hard-to-read or understand. Specifically, regarding the amount of presented information, 1 participant mentioned that too much information was presented, 2 participants mentioned a cognitive overload by the number of films/charts presented, and 2 participants also mentioned that they needed some time to understand the graphs. Regarding the design of the visualization, 2 participants stated that the doughnut/pie charts were not appropriate in this context, 3 stated that the encasing of the rings for the gender and age was confusing for them, 3 mentioned that the color coding was confusing (e.g., $\bullet$ ``it would be easier to understand if there were different colors for different categories, like Male over 50s is a different color than Female over 50s'' $\bullet$ ``green represents male and female aged over 50 at the same time''), 2 did not like hovering to see the labels (another participant found this feature useful). 

Another source of confusion mentioned by 3 participants was what the percentages represented/how they were calculated (e.g., $\bullet$ ``I am confused with how the percentages are calculated, whether it is counting the cast or measuring screentime.'' $\bullet$ ``[...] like if a character leaves the scene and then comes back does that count as two?''). Finally, 2 participants misunderstood (elements of) the design (e.g., $\bullet$ ``There was no information on male appearances'' $\bullet$ ``the description of male or females over 50 should’ve been specified'').

Regarding the interpretability of the tool, 2 participants found the info icons with the popups explaining what confidence and bias mean useful, while 1 participant would like the shown texts to be simpler and more descriptive. 4 participants found the bias bar graphs useful, while 2 participants referred to the models as being accurate, although we did not provide information about their accuracy. A participant mentioned they doubt that ``age can be discerned visually by AI given how rampant plastic surgery is in most global film industries [...] feels like it shouldn't be possible''. 2 participants mentioned they would like to know the films shown in the graphs, and one of them said ``I don't know what films these are based on for me to double-check - that's why I'm not very confident in the AI's judgement (looking at the bottom charts - the AI is very close to the ground truth so that increases my confidence now).''

Five participants mentioned that the UI could be improved and some participants (4) provided certain suggestions like using bar graphs instead of doughnut charts, a different color for each age/gender group and a legend, confidence interval in the doughnut chart, icons to represent gender/age groups. 2 participants also raised a point for the demographics to be extended and include non-binary gender categories and ethnicity/race.

Only one participant wrote: ``For a movie that I am intending to watch I would rather read reviews about how the characters are presented and portrayed. Quantitative measures feel meaningless to me in this circumstances, the numbers do not mean any specific thing. I am interested in what a movie presents and with what qualities rather than any measurements''.
\vspace{-3mm}
\subsection{Discussion}
\vspace{1mm}
\noindent\textbf{Summary of findings}.
Our character representation and visualization tool for media content presented the information at a satisfactory level. Most participants were able to understand the presented information for gender, age and their intersection (Q2.2-3,5-6) and the model-related information (Q2.7-8) quite accurately. There is quite some uncertainty around people's understanding of what the percentages represented (Q2.1), which was also raised by 3 participants in Q3.10. The accuracy of responses in Q2.4 is also uncertain but this might be due to the broad nature of the question.

Participants' trust in our models was moderate for the gender predictions, while below the moderate level for age. There could be various reasons for this. AI skepticism could be one reason. One participant openly expressed their skepticism for AI in Q3.11 but the use of AI in this context was deemed moderately useful in Q3.3. The low confidence of people in the age model could be attributed to their doubts about the ability of AI to recover the true age of faces given the prevalence of appearance interventions (one participant raised this point in Q3.10), the varying AI confidence for the age prediction across the 3 movies, and the slight bias for characters over 50 might have challenged people's trust. 4 participants found the bias bar graphs very useful in Q3.10 and the usefulness of the AI confidence and bias information was rated above average (Q3.4-5), but we do not know if this information might have caused a reverse effect on their trust in alignment with previous work \cite{Kizilcec2016,Papenmeier2019}.
Nevertheless, there were 2 participants, who deemed our models accurate based on this information.

The usefulness of doughnut charts in this context was deemed mediocre (Q3.1). 2 participants in Q3.10 clearly stated that they did not believe these graphs were very appropriate. This could be partly attributed to some design choices that apparently confused some participants; 3 participants in Q3.10 found the color coding/nesting of rings confusing, and 2 did not like hovering to see the labels. 

The fact that we presented 3 doughnut charts for 3 different films at the same time along with the complexity of the nested information and having to hover to reveal the labels at the intersection of gender and age probably made some people to perceive this interface as confusing or state that they needed some time to understand the graphs (Q3.10). A clearer experimental design would have been helpful here.

Participants also deemed the on-screen appearances distribution at the intersection of gender and age (Q3.2) moderately useful. This could be either because participants are not interested in this information (4 participants in Q3.11 stated their lack of interest about the presented information) or the way this distribution was presented was too confusing for people (3 participants in Q3.10 found the nested rings confusing). The cognitive load was moderate (Q3.6), the frustration low (Q3.7), and the information understandability at the intersection of gender and age quite accurate (Q2.6), while only 13.33\% of participants self-reported their age as greater than 40. It is also noteworthy here that 5 participants mentioned that would be interested in more extended demographics, mainly ethnicity/race. 4 participants also discussed their belief that more context around the portrayals of characters is required and quantitative comparisons might not be that useful for the audience. One third of the participants stated that they would not use our tool, or they are not sure about its applicability.

The participants' liking of using our tool was moderate (Q3.9), while the overall rating of our tool (Q3.8) was above average. These findings seem to suggest that our tool is still relevant, interesting, and useful. 40\% of the participants think that the existing streaming platforms/movie database sites are slightly or not at all informative about the diversity/inclusivity of characters in films and tv series (Q1.6), while some identified a broader applicability of such tools in media more broadly, and beyond (Q3.11). Nonetheless, based on the data we collected, there is scope for 
\begin{itemize}    
    \item design improvements in the presentation of the information especially how the intersection of demographics is presented and where the color coding puts the emphasis on,
    \item including more demographics, like ethnicity/race, and more contextual analyzes of the presented characters,
    \item accounting for more tailored visualizations depending on the interests of the audience, e.g. by adding more interactivity and configurability, and
    \item increasing the trustworthiness of the tool e.g., by providing more/better explanations to disentangle concepts like accuracy, bias and confidence, clarify the limitations of the models (e.g. only the \emph{perceived} gender and age can be predicted).
\end{itemize}

\vspace{1mm}
\noindent\textbf{Future directions.}
Our tool’s development is characterized by modularity and input standardisation, which make it easily extendable to include e.g., other demographic dimensions, more context or interactivity. Due to the lack of previous studies in this topic, we kept the design of the study as simple as possible to avoid overwhelming people and allowing space to ask about various other aspects of the tool than its design, such as trust. Comparative studies are required to test the effects of alternative (refined) designs of the visualization or the effects of AI- vs. human-generated analytics. While we started with a simple design and study, those could be future steps in this work.

We believe that AI and visualization are valuable tools for analyzing the character representation in media and communicating these analyzes to the broader audience. In the era of Over-the-top production of media content, tools that could automate such analyzes would help people better understand the offered content and identify what is relevant and interesting to them through a perspective not currently available to them. But there is need for more work to better map the audience's expectations from such tools, and strike a balance in the complex interplay of explainability, accuracy, and trust \cite{Papenmeier2022}.

\section{Conclusion}
Recent advances in AI has made it possible to analyze media content with reasonable accuracy at scale and generate character representation analytics. Getting the audience in the loop within this process is important. We need to understand what matters to them and whether they would trust such an analysis system. We built an open-source tool, which is not an off-the-shelf AI facial recognition system but is a frontier multimodal model employed in a novel way: based on the CLIP model we analyze the character representation across dimensions of age and gender and visualize the extracted analytics. We conducted a user study to provide empirical evidence on the usefulness and trustworthiness of the AI-generated results for three full length movies presented in form of our visualizations. Participants understood well the presented information and deemed the tool overall useful but there is a request for more tailored visualizations that include more demographic categories and contextual analyzes of the character portrayals. Participants did not reject the use of AI in this context, they found the presented information about the AI confidence and bias very useful, but their trust in our gender and age models was mediocre and low, respectively. These are just some initial findings but more work is required towards this direction. In the era of over-the-top production of media content, tools like this one will become more and more important in helping people understand the film through the lens of film consumption and decision-making about cultural or social content.

\section*{Safe and Responsible Innovation Statement}
The benchmarking of the models was performed on a free and publicly  available dataset. We legally acquired the video content we analyzed. The user study was approved by relevant authority at the authors' institution. All participants data collected is anonymized and therefore individuals cannot be personally identified.  We carefully considered the risks for participants in taking part in the user study. We minimized the risk of participant fatigue by setting a maximum time limit of 30 minutes. Our tool should be only used with legally acquired media content. The intellectual property of creatives, fairness and consent are key issues our research is committed to.
\section*{Acknowledgment}
This work was funded by the Leverhulme Trust (RPG-2023-091).
\balance
\bibliographystyle{ACM-Reference-Format}
\bibliography{sample-base}
\appendix

\end{document}